\documentclass{ws-mpla}

\usepackage{xspace}
\def\deltam     {\ensuremath{\Delta m}\xspace}

\def\DDbar   {\ensuremath{\kern -0.1em \stackrel{\kern 0.1em \textsf{\fontsize{5pt}{1em}\selectfont(---)}}{D}\kern -0.3em}\xspace}
\def\AAbar   {\ensuremath{\kern -0.2em \stackrel{\kern 0.2em \textsf{\fontsize{5pt}{1em}\selectfont(---)}}{A}\kern -0.3em}\xspace}
\def\AfAfbar   {\ensuremath{\kern -0.2em \stackrel{\kern 0.2em \textsf{\fontsize{5pt}{1em}\selectfont(---)}}{A}_{\kern -0.3em f}\kern -0.3em}\xspace}
\def\dacp     {\ensuremath{\Delta A_{\CP}}\xspace}
\def\CP                {\ensuremath{C\!P}\xspace}
\def\Dbar    {\kern 0.2em\overline{\kern -0.2em \PD}{}\xspace}
\def\D       {\ensuremath{\PD}\xspace}
\def\Db      {\ensuremath{\Dbar}\xspace}
\def\Dz      {\ensuremath{\D^0}\xspace}
\def\Dzb     {\ensuremath{\Dbar^0}\xspace}
\def\DzDzb   {\ensuremath{\Dz {\kern -0.16em \Dzb}}\xspace}
\def\Dp      {\ensuremath{\D^+}\xspace}
\def\Dm      {\ensuremath{\D^-}\xspace}

\def\DpDm    {\ensuremath{\Dp {\kern -0.16em \Dm}}\xspace}
\def\Dstar   {\ensuremath{\D^*}\xspace}

\def\Dstarp  {\ensuremath{\D^{*+}}\xspace}

\def\DsJ     {\ensuremath{\D_{\squark J}}\xspace}
\def\Kstar   {\ensuremath{\PK^*}\xspace}
\def\lhcb {LHCb\xspace}

\def\babar  {BaBar\xspace}
\def\belle  {Belle\xspace}
\def\cleo   {CLEO\xspace}

\def\cdf    {CDF\xspace}
\def\PD      {\ensuremath{D}\xspace}                 
\def\PM      {\ensuremath{M}\xspace}                 
\def\PB      {\ensuremath{B}\xspace}                 
\def\Bz      {\ensuremath{\B^0}\xspace}
\def\Bzb     {\ensuremath{\Bbar^0}\xspace}
\def\Bp      {\ensuremath{\B^+}\xspace}
\def\Bm      {\ensuremath{\B^-}\xspace}
\def\PK      {\ensuremath{K}\xspace}                 
\def\kaon  {\ensuremath{\PK}\xspace}
\def\Kp    {\ensuremath{\kaon^+}\xspace}
\def\Km    {\ensuremath{\kaon^-}\xspace}
\def\Ppi         {\ensuremath{\pi}\xspace}                 
\def\Prho        {\ensuremath{\rho}\xspace}                 
\def\pion  {\ensuremath{\Ppi}\xspace}
\def\pip   {\ensuremath{\pion^+}\xspace}
\def\pim   {\ensuremath{\pion^-}\xspace}
\def\ycp        {\ensuremath{y_{CP}}\xspace}
\def\agamma     {\ensuremath{A_{\Gamma}}\xspace}
\def\KS    {\ensuremath{\kaon^0_{\rm\scriptstyle S}}\xspace} 
\def\KL    {\ensuremath{\kaon^0_{\rm\scriptstyle L}}\xspace} 
\newcommand{\ie}{\mbox{\itshape i.e.}}
\newcommand{\eg}{\mbox{\itshape e.g.}}
\def\Bbar    {\kern 0.2em\overline{\kern -0.2em \PB}{}\xspace}
\def\B       {\ensuremath{\PB}\xspace}

\def\Jpsi     {\ensuremath{{\PJ\mskip -3mu/\mskip -2mu\Ppsi\mskip 2mu}}\xspace}
\def\PJ      {\ensuremath{\mathrm{J}}\xspace}                 
\def\Ppsi        {\ensuremath{\psi}\xspace}                 
\def\Pupsilon    {\ensuremath{\Upsilon}\xspace}                 
\def\squark    {\ensuremath{\Ps}\xspace}
\def\Ps      {\ensuremath{\mathrm{s}}\xspace}                 
\def\dquark    {\ensuremath{\Pd}\xspace}
\def\Pd      {\ensuremath{\mathrm{d}}\xspace}                 

\def\Pu      {\ensuremath{\mathrm{u}}\xspace}                 
\def\cquark    {\ensuremath{\Pc}\xspace}
\def\Pc      {\ensuremath{\mathrm{c}}\xspace}                 
\def\ubarquark    {\ensuremath{\overline{\Pu}}\xspace}
\def\cbarquark    {\ensuremath{\overline{\Pc}}\xspace}

\newcommand{\decay}[2]{\ensuremath{#1\!\to #2}\xspace}         
\def\DDbar   {\ensuremath{\kern -0.1em \stackrel{\kern 0.1em \textsf{\fontsize{5pt}{1em}\selectfont(---)}}{D}\kern -0.3em}\xspace}
\def\AAbar   {\ensuremath{\kern -0.2em \stackrel{\kern 0.2em \textsf{\fontsize{5pt}{1em}\selectfont(---)}}{A}\kern -0.3em}\xspace}
\def\AfAfbar   {\ensuremath{\kern -0.2em \stackrel{\kern 0.2em \textsf{\fontsize{5pt}{1em}\selectfont(---)}}{A}_{\kern -0.3em f}\kern -0.3em}\xspace}
\def\dacp     {\ensuremath{\Delta A_{\CP}}\xspace}
\def\deltam   {\ensuremath{\delta m}\xspace}
\def\besiii {BESIII\xspace}
\def\Mbar    {\kern 0.2em\overline{\kern -0.2em \PM}{}\xspace}
\def\M       {\ensuremath{\PM}\xspace}
\def\Mz      {\ensuremath{\M^0}\xspace}
\def\Mzb     {\ensuremath{\Mbar^0}\xspace}
\def\Pe      {\ensuremath{\mathrm{e}}\xspace}                 
\def\ep         {\ensuremath{\Pe^+}\xspace}
\def\en         {\ensuremath{\Pe^-}\xspace}
\def\epem       {\ensuremath{\Pe^+\Pe^-}\xspace}
\def\mup        {\ensuremath{\Pmu^+}\xspace} 
\def\mun        {\ensuremath{\Pmu^-}\xspace} 
\def\mump       {\ensuremath{\Pmu^\mp}\xspace} 
\def\epm       {\ensuremath{\Pe^\pm}\xspace} 
\def\Pmu         {\ensuremath{\mu}\xspace}                 
\def\neu        {\ensuremath{\Pnu}\xspace}
\def\neub       {\ensuremath{\overline{\Pnu}}\xspace}

\def\neum       {\ensuremath{\neu_\mu}\xspace}
\def\neumb      {\ensuremath{\neub_\mu}\xspace}

\def\Pnu         {\ensuremath{\nu}\xspace}                 

\def\mup        {\ensuremath{\Pmu^+}\xspace}
\def\pion  {\ensuremath{\Ppi}\xspace}
\def\piz   {\ensuremath{\pion^0}\xspace}

\def\pip   {\ensuremath{\pion^+}\xspace}
\def\pim   {\ensuremath{\pion^-}\xspace}
\def\lhcb {LHCb\xspace}
\def\ux85 {UX85\xspace}

\def\babar  {BaBar\xspace}
\def\belle  {Belle\xspace}

\def\cdf    {CDF\xspace}

\def\cleo   {CLEO\xspace}

\def\tevatron {Tevatron\xspace}
 \def\Pphi        {\ensuremath{\phi}\xspace}                 
  \def\Dbar    {\kern 0.2em\overline{\kern -0.2em \PD}{}\xspace}
\def\D       {\ensuremath{\PD}\xspace}
\def\Db      {\ensuremath{\Dbar}\xspace}
\def\Dz      {\ensuremath{\D^0}\xspace}
\def\Dzb     {\ensuremath{\Dbar^0}\xspace}
\def\DzDzb   {\ensuremath{\Dz {\kern -0.16em \Dzb}}\xspace}
\def\Dp      {\ensuremath{\D^+}\xspace}
\def\Dm      {\ensuremath{\D^-}\xspace}

\def\DpDm    {\ensuremath{\Dp {\kern -0.16em \Dm}}\xspace}
\def\Dstar   {\ensuremath{\D^*}\xspace}

\def\Dstarp  {\ensuremath{\D^{*+}}\xspace}

\def\Ds      {\ensuremath{\D^+_\squark}\xspace}
\def\Dsp     {\ensuremath{\D^+_\squark}\xspace}
\def\Dsm     {\ensuremath{\D^-_\squark}\xspace}

\def\Bd      {\ensuremath{\B^0_\dquark}\xspace}
\def\Bs      {\ensuremath{\B^0_\squark}\xspace}

\def\Wpm    {\ensuremath{\PW^\pm}\xspace}
\def\PW      {\ensuremath{W}\xspace}

\def\mbarn{\ensuremath{\rm \,mb}\xspace}
\def\mub{\ensuremath{\rm \,\mu b}\xspace}

\def\nb {\ensuremath{\rm \,nb}\xspace}

\def\invfb   {\ensuremath{\mbox{\,fb}^{-1}}\xspace}

\newcommand{\tev}{\ensuremath{\mathrm{\,Te\kern -0.1em V}}\xspace}
\newcommand{\gev}{\ensuremath{\mathrm{\,Ge\kern -0.1em V}}\xspace}
\newcommand{\mev}{\ensuremath{\mathrm{\,Me\kern -0.1em V}}\xspace}
\newcommand{\kev}{\ensuremath{\mathrm{\,ke\kern -0.1em V}}\xspace}
\newcommand{\ev}{\ensuremath{\mathrm{\,e\kern -0.1em V}}\xspace}
\newcommand{\gevc}{\ensuremath{{\mathrm{\,Ge\kern -0.1em V\!/}c}}\xspace}
\newcommand{\mevc}{\ensuremath{{\mathrm{\,Me\kern -0.1em V\!/}c}}\xspace}
\newcommand{\gevcc}{\ensuremath{{\mathrm{\,Ge\kern -0.1em V\!/}c^2}}\xspace}
\newcommand{\gevgevcccc}{\ensuremath{{\mathrm{\,Ge\kern -0.1em V^2\!/}c^4}}\xspace}
\newcommand{\mevcc}{\ensuremath{{\mathrm{\,Me\kern -0.1em V\!/}c^2}}\xspace}

\usepackage{ifthen} 

\begin{document}

\newboolean{articletitles}
\setboolean{articletitles}{true} 

\markboth{Marco Gersabeck}
{Brief Review of Charm Physics}

\catchline{}{}{}{}{}

\title{BRIEF REVIEW OF CHARM PHYSICS}

\author{\footnotesize Marco Gersabeck}

\address{CERN, 1211 Geneva, Switzerland\\
marco.gersabeck@cern.ch}

\maketitle

\pub{Received (Day Month Year)}{Revised (Day Month Year)}

\begin{abstract}
Charm physics has attracted increased attention after first evidence for charm mixing was observed in 2007.
The level of attention has risen sharply after \lhcb reported first evidence for \CP violation in the charm sector.
Neither mixing nor \CP violation have been established by a single unambiguous measurement to date.
This review covers the status of mixing and \CP violation measurements and comments on the challenges on the road ahead, both on the experimental and theoretical side,  and on ways to tackle them.

\end{abstract}

\ccode{PACS Nos.: 13.20.Fc, 13.25.Ft, 14.40.Lb}

\section{Introduction}

Charm physics covers the studies of a range of composite particles containing charm quarks which provide unique opportunities for probing the strong and weak interactions in the standard model and beyond.
The charm quark, being the up-type quark of the second of the three generations, is the third-heaviest of the six quarks.
Charm particles can exist as so-called open charm mesons or baryons, containing one or several (for baryons) charm quarks, or as charmonium states which are bound states of charm-anticharm quark pairs.

The uniqueness of charm particles lies in their decays.
The charm quark can only decay via annihilation with an anti-charm quark in the case of charmonium states or as a weak decay, mediated by a \Wpm-boson, into a strange or down quark.
Thus, open charm particles are the only ones allowing the study of weak decays of an up-type quark in a bound state.

In 2009, Ikaros Bigi asked whether charm's third time could be the real charm~\cite{Bigi:2009jj}.
Charm's first time was the discovery of the \Jpsi~\cite{Aubert:1974js,Augustin:1974xw}, which followed three years after the possible first observation of an open charm decay in cosmic ray showers~\cite{Niu:1971xu}. 
This discovery confirmed the existence of a fourth quark as expected by the GIM mechanism~\cite{Glashow:1970gm} motivated by the non-existence of flavour-changing neutral currents~\cite{Bott:1967,Foeth:1969hi} in conjunction with the observation of the mixing of neutral kaons~\cite{Lande:1956pf,Jackson:1957zzb,Niebergall:1974wh}.
The second time charm attracted considerable attention was caused by the observation of \DsJ states~\cite{Aubert:2003fg,Besson:2003cp,Abe:2003jk,Aubert:2003pe} which could not be accommodated by QCD~\cite{Godfrey:1985xj,Godfrey:1986wj,Isgur:1991wq,DiPierro:2001uu,Matsuki:2007zza}.
Until today, excited charmonium and open charm particles provide an excellent laboratory for studying QCD, however, this topic is beyond the scope of this review.

Charm's third time started with the first evidence for the mixing of neutral charm mesons reported by \babar~\cite{Aubert:2007wf} and \belle~\cite{Staric:2007dt} in 2007.
Since then a lot of work went into more precise measurements of the mixing phenomenon as well as into searches for charge-parity (\CP) symmetry violation in the charm sector.
At the same time theoretical calculations were improved even though precise standard model predictions are still a major challenge.
This paper reviews the current situation of studies of processes, which are mediated by the weak interaction at leading order, using open charm particles.
Particular focus is given to mixing and \CP violation, followed by comments on rare charm decays at the end of this review.

\subsection{Charm production}
\label{sec:charm_prod}

Charm physics has been and is being performed at a range of different accelerators.
These come with different production mechanisms and thus with largely varying production cross-sections.
At $\ep\en$ colliders two different running conditions are of interest to charm physics.
Tuning the centre-of-mass energy to resonantly produce $\Ppsi(3770)$ states leads to the production of quantum-correlated $\Dz\Dzb$ or $\Dp\Dm$ pairs.
This is the case for the CLEO-c experiment at the CESR-c collider as well as for BESIII at BEPCII.
The most commonly used alternative is running at a higher centre-of-mass energy to resonantly produce $\Pupsilon(4S)$ which decay into quantum-correlated $\Bz\Bzb$ or $\Bp\Bm$ pairs.
This is used by the \babar and \belle experiments which are located at the PEP-II and KEKB colliders, respectively.
Both PEP-II and KEKB are asymmetric colliders, thus having a collision system that is boosted with respect to the laboratory frame.
This allows measurements with decay-time resolutions about a factor two to four below the \Dz lifetime and therefore decay-time dependent studies.

The production cross-section for producing $\D\Db$ pairs at the $\Ppsi(3770)$ resonance is approximately $8\nb$~\cite{Ablikim:2004ck}.
When running at the $\Pupsilon(4S)$ resonance, the cross-section for producing $\cquark\cbarquark$ pairs is $1.3\nb$~\cite{Harrison:1998yr}.
The latter scenario gives access to all species of charm particles while the $\Ppsi(3770)$ only decays into $\Dz\Dzb$ or $\Dp\Dm$ pairs.
The \babar and \belle experiments have collected integrated luminosities of about $500\invfb$ and $1000\invfb$, respectively.
CLEO-c has collected $0.5\invfb$ at the $\Ppsi(3770)$ resonance as well as around $0.3\invfb$ above the threshold for $\Dsp\Dsm$ production.
BESIII has so far collected nearly $3\invfb$ in their 2010 and 2011 runs.

At hadron colliders the production cross-sections are significantly higher.
The cross-section for producing $\cquark\cbarquark$ pairs in proton-proton collisions at the LHC with a centre-of-mass energy $7\tev$ is about $6\mbarn$~\cite{LHCb-CONF-2010-013}, \ie\ more than six orders of magnitude higher compared to operating an $\ep\en$ collider at the $\Pupsilon(4S)$ resonance.
This corresponds to a cross-section of about $1.5\mbarn$ for producing $\Dz$ in the \lhcb acceptance\footnote{The \lhcb acceptance is given as a range in momentum transverse to the beam direction and rapidity as $p_{\rm T}<8\gevc,2<y<4.5$.}.
This number may be compared to its equivalent at \cdf which has been measured to $13\mub$ inside the detector acceptance\footnote{The \cdf acceptance is defined as $p_{\rm T}>5.5\gevc,|y|<1$.} for proton-antiproton collisions at the \tevatron with $\sqrt{s}=1.96\tev$~\cite{Acosta:2003ax}.
\cdf has collected a total of about $10\invfb$ while \lhcb has collected about $1.8\invfb$ by the time of writing this review, corresponding to $1.3\times10^{11}$ and $2.7\times10^{12}$ \Dz mesons produced in the respective detector acceptances.

The production of charm quarks in hadron collisions occurs predominantly in very asymmetric collisions which result in heavily boosted quarks with high rapidities.
Therefore, \lhcb is ideally suited for performing decay-time dependent studies of charm decays.
At the same time, the $\cquark\cbarquark$ cross-section at the LHC is about $10\%$ of the total inelastic cross-section which allows to have reasonably low background levels for a hadronic environment.
The coverage of nearly the full solid angle of the $\ep\en$-collider experiments mentioned here makes them very powerful instruments for analysing decays involving neutral particles that may remain undetected or for inclusive studies.

\subsection{Mixing and \CP violation in neutral mesons}
\label{sec:cpv_intro}

For neutral mesons, the mass eigenstates, \ie\ the physical particles, generally do not coincide with the flavour eigenstates, \ie\ those governing the interactions.
The mass eigenstates of neutral mesons, $|\PM_{1,2}\rangle$, with masses $m_{1,2}$ and widths $\Gamma_{1,2}$, are linear
combinations of the flavour eigenstates, $|\Mz\rangle$ and $|\Mzb\rangle$, as $|\PM_{1,2}\rangle=p|\Mz\rangle\pm{}q|\Mzb\rangle$ with
complex coefficients satisfying $|p|^2+|q|^2=1$.
This allows the definition of the averages $m\equiv(m_1+m_2)/2$ and $\Gamma\equiv(\Gamma_1+\Gamma_2)/2$.
The phase convention of $p$ and $q$ is chosen such that, in the limit of no \CP violation, $\CP|\Mz\rangle=-|\Mzb\rangle$.

Mixing, \ie\ the periodical transformation of mesons into their anti-mesons and back, occurs if there is a non-zero difference in the masses or widths of the two mass eigenstates of a meson.
This is quantified in the differences $\Delta{}m\equiv{}m_2-m_1$ and $\Delta\Gamma\equiv\Gamma_2-\Gamma_1$.
Furthermore, the mixing parameters are defined as $x\equiv\Delta{}m/\Gamma$ and $y\equiv\Delta\Gamma/(2\Gamma)$.

It is these mixing parameters $x$, and $y$ which define the characteristic behaviour of the four neutral meson systems, which are subject to mixing, kaons (\PK), charm (\PD), \Bd, and \Bs mesons.
To appreciate the different mixing behaviour it is instructive to consider the time evolution of the meson and anti-meson states.
The probability of observing a neutral meson state \Mz or \Mzb after a time $t$ has passed since the observation of an initial state \Mz is
\begin{align}
\label{eqn:charm_start}
P(\Mz\to\Mz,t)&=\frac{1}{2}e^{-\Gamma{}t}(\cosh(y\Gamma{}t)+\cos(x\Gamma{}t)),\nonumber\\*
P(\Mz\to\Mzb,t)&=\frac{1}{2}\left|\frac{q}{p}\right|^{2}e^{-\Gamma{}t}(\cosh(y\Gamma{}t)-\cos(x\Gamma{}t)),
\end{align}
where the oscillatory behaviour is governed by the mixing parameter $x$.

\begin{figure}
\centering
\includegraphics[width=1.0\textwidth]{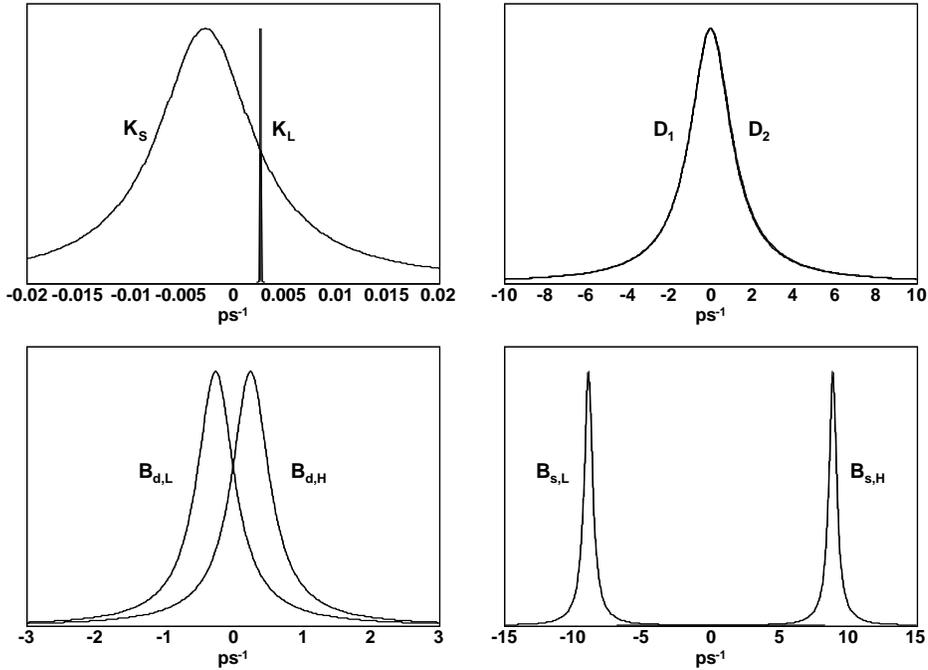}
\caption{The widths and mass differences of the physical states of the flavoured neutral mesons. The width corresponds to the inverse lifetime while the mass difference determines the oscillation frequency.}
\label{fig:mesons}
\end{figure}

For charm mesons the mixing parameters are drastically different compared to those of kaons or \PB mesons.
Figure~\ref{fig:mesons} shows the widths and mass differences of the four neutral meson systems.
The kaon system is the only one to have $y\approx 1$, resulting in two mass eigenstates with vastly different lifetimes, hence their names \PK-short (\KS) and \PK-long (\KL).
Furthermore, also $x\approx1$ which results in a sizeable sinusoidal oscillation frequency as shown in Eq.~(\ref{eqn:charm_start}).
The two \PB-meson systems have reasonably small width splitting, however, they have sizeable values for $x$.
Particularly for the \Bs system this leads to fast oscillations which require high experimental accuracy to be resolved.
The charm meson system is the only one where both $x$ and $y$ are significantly less than $1$, hence the nearly overlapping curves in Fig.~\ref{fig:mesons}.

Experimentally, the different mixing parameters lead to rather different challenges for measurements in the various meson systems.
The vast lifetime difference in the kaon system leads to the possibility of studying nearly clean samples of just one of the two mass eigenstates by either measuring decays close to a production target where \KS decays dominate, or far away where most \KS have decayed before entering the detection region.
In the \PB systems the oscillation frequency puts a challenge to the decay-time resolution, particularly for \Bs mesons as mentioned before.
The smallness of $y$ requires, to first order, large data samples to acquire the necessary statistical precision for measuring such a small quantity.
The latter is particularly true for the charm system, where both $x$ and $y$ are small.
This is the reason why it was only in 2007 when first evidence for charm mixing was observed.

The symmetry under \CP transformation is violated for a deviation from unity of the quantity $\lambda_f$, defined as
\begin{equation}
\label{eqn:charm_lambda}
\lambda_f\equiv\frac{q\bar{A}_{\bar{f}}}{pA_f}=-\eta_{\CP}\left|\frac{q}{p}\right|\left|\frac{\bar{A}_f}{A_f}\right|e^{i\phi},
\end{equation}
where the right-hand expression is valid for a \CP eigenstate $f$ with eigenvalue $\eta_{\CP}$ and $\phi$ is the \CP violating relative phase between $q/p$ and $\bar{A}_f/A_f$.
Besides the mixing parameters introduced above this expression contains the decay amplitudes $A_f$ and $\bar{A}_f$ for decays into a final state $f$.

\CP violation can have different origins: the case $|q/p|\neq1$ is called \CP violation in mixing, $|\bar{A}_f/A_f|\neq1$ is \CP violation in
the decay, and a non-zero phase $\phi$ between $q/p$ and $\bar{A}_f/A_f$ causes \CP violation in the interference between mixing and decay.
Mixing is common to all decay modes and hence \CP violation originating in this process is universal which is called indirect \CP violation.
Decay-specific \CP violation is called direct \CP violation.
An excellent discussion on the different types of \CP violation can be found in section 7.2.1 of Ref.~\refcite{Sozzi:2008zza}.
As opposed to the strange and the beauty system, \CP violation has not yet been discovered in the charm system,
though the \lhcb collaboration has recently found first evidence for \CP violation in two-body \Dz decays~\cite{Aaij:2011in}, arguably the most surprising result from the LHC in 2011.

\section{Charm mixing}
\label{sec:mix_charm}
The studies of charm mesons have gained in momentum with the measurements of first evidence for meson anti-meson mixing in neutral charm mesons in 2007~\cite{Aubert:2007wf,Staric:2007dt}.
Mixing of \Dz mesons is the only mixing process where down-type quarks contribute to the box diagram.
Unlike \PB-meson mixing, where the top-quark contribution dominates, the third generation quark is of similar mass to the other down-type quarks.
This leads to a combination of GIM cancellation~\cite{Glashow:1970gm} and CKM suppression~\cite{Cabibbo:1963yz,Kobayashi:1973fv}, which results in a strongly suppressed mixing process~\cite{Georgi:1992as,Ohl:1992sr,Bigi:2000wn,Chen:2007zua,Bobrowski:2010xg}.

There are two approaches for theoretical calculations of charm mixing.
The ``inclusive'' approach is based an operator product expansion (OPE) in $\Lambda/m_{\cquark}$~\cite{Khoze:1983yp,Shifman:1984wx,Chay:1990da,Bigi:1992su,Blok:1993va,Manohar:1993qn,Bobrowski:2010xg}.
Due to the cancellations mentioned above it is higher order operators that give the largest contributions to the mixing parameters.
Furthermore, it is not yet clear whether the expansion series really converges.
Calculations of the charm meson lifetimes are being performed to test whether the OPE approach can properly reproduce the large difference between the \Dz and the \Dp lifetimes.
In the \Bs system, the OPE approach successfully predicted the width splitting of the two \Bs mass eigenstates~\cite{Lenz:2011ti} which has recently been confirmed by an \lhcb measurement~\cite{LHCb-CONF-2012-002}.

The ``exclusive'' approach sums over intermediate hadronic states, taking input from models or experimental data~\cite{Donoghue:1985hh,Wolfenstein:1985ft,Colangelo:1990hj,Kaeding:1995zx,Anselm:1979fa,Cheng:2010rv}.
Also in this approach, different modes of the same $SU(3)$ multiplet lead to cancellations which is why their individual contributions have to be known to high precision.
Due to the considerable mass of the \Dz meson, many different modes need to be taken into account simultaneously.
Of these, only phase space differences can be evaluated at the moment.
Estimates indicate that mixing in the experimentally observed range is conceivable when taking into account $SU(3)$-breaking effects.
However, neither the inclusive nor the exclusive approach have thus far permitted a precise theoretical calculation of charm mixing.

It was discussed whether the measured size of the mixing parameters could be interpreted as a hint for physics beyond the standard model~\cite{Hou:2006mx,Ciuchini:2007cw,Nir:2007ac,Blanke:2007ee,He:2007iu,Chen:2007yn,Golowich:2007ka}.
The biggest problem in answering this question is the non-existence of a precise standard model calculation.
Effects of physics beyond the standard model were also searched for in numerous \CP violation measurements and searches for rare decays both of which are covered in the remainder of this review.

Mixing of \Dz mesons can be measured in several different modes.
Most require identifying the flavour of the \Dz at production as well as at the time of the decay.
Tagging the flavour at production usually exploits the strong decay \decay{\Dstarp}{\Dz\pip} (and charge conjugate\footnote{Charge conjugate decays are implicitly included henceforth.}) where the charge of the pion determines the flavour of the \Dz.
The small amount of free energy in this decay leads to the difference in the reconstructed invariant mass of the \Dstarp and the \Dz, $\deltam\equiv{}m_{\Dstarp}-m_{\Dz}$, exhibiting a sharply peaking structure over a threshold function as background.
An alternative to using this decay mode is tagging the \Dz flavour by reconstructing a flavour-specific decay of a \PB meson.
This method has not yet been used in a measurement as it did not yet yield competitive quantities of tagged \Dz mesons.
At \lhcb this approach may be of interest due to differences in trigger efficiencies compensating for lower production rates.
Another option available particularly at $\epem$ colliders is the reconstruction of the opposite side charm meson in a flavour specific decay.

Theoretically, the most straight-forward mixing measurement is that of the rate of the forbidden decay \decay{\Dz}{\Kp\mun\neumb} which is only accessible through \Dz-\Dzb mixing.
The ratio of the time-integrated rate of these forbidden decays to their allowed counterparts, \decay{\Dz}{\Km\mup\neum}, determines $R_{\rm m}\equiv(x^2+y^2)/2$.
As this requires very large samples of \Dz mesons no measurement has thus far reached sufficient sensitivity to see evidence for \Dz mixing.
The most sensitive measurement to date has been made by the \belle collaboration~\cite{Bitenc:2008bk} to $R_{\rm m}=(1.3\pm2.2\pm2.0)\times10^{-4}$, where the first uncertainty is of statistical and the second is of systematic nature\footnote{This notation is applied to all results where two uncertainties are quoted.}.

Related to the semileptonic decay is the suppressed decay \decay{\Dz}{\Kp\pim}, called wrong-sign (WS) decay.
For this decay, a doubly Cabibbo-suppressed (DCS) amplitude interferes with the decay through a mixing process followed by the Cabibbo-favoured (CF) decay \decay{\Dz}{\Km\pip}.
The time-dependent decay rate of the WS decay is, in the limit of \CP conservation, proportional to
\begin{equation}
\frac{\Gamma(\Dz(t)\to \Kp\pim)}{e^{-\Gamma t}}\propto \left(R_{\rm D}+\sqrt{R_{\rm D}}y'\Gamma{}t+R^2_{\rm m}(\Gamma{}t)^2\right),
\end{equation}
where the mixing parameters are rotated by the strong phase between the DCS and the CF amplitude, leading to the observable $y'=y\cos\delta_{\PK\Ppi}-x\sin\delta_{\PK\Ppi}$~\cite{Bergmann:2000id}.
The parameter $R_{\rm D}$ is the ratio of the DCS to the CF rate.
Measurements with sufficient sensitivity to unveil evidence for \Dz mixing have been performed by the \babar and \cdf collaborations, leading to
\begin{table}[h!!!]
\centering
\begin{tabular}{lcc}
           & $x'^2$ in $10^{-3}$ & $y'$ in $10^{-3}$\\
\hline
\babar~\cite{Aubert:2007wf} & $-0.22\pm0.30\pm0.20$ & $9.7\pm4.4\pm3.1$\\
\cdf~\cite{Aaltonen:2007uc} & $-0.12\pm0.35$ & $8.5\pm7.6$.
\end{tabular}
\end{table}

Similarly, the CF and DCS amplitudes can also lead to excited states of the same quark content.
The decay \decay{\Dz}{\Km\pip\piz} is the final state of several such resonances.
Thus, by studying the decay-time dependence of the various resonances a mixing measurement can be obtained.
The \babar collaboration achieved a measurement showing evidence for \Dz mixing~\cite{Aubert:2008zh} with central values of $x''=(26.1^{+5.7}_{-6.8}\pm3.9)\times10^{-3}$ and $y''=(-0.6^{+5.5}_{-6.4}\pm3.4)\times10^{-3}$, where the rotation between the observables and the system of mixing parameters is given by a strong phase as 
\begin{align}
x''&=x\cos\delta_{\Km\pip\piz}+y\sin\delta_{\Km\pip\piz}\\*
y''&=y\cos\delta_{\Km\pip\piz}-x\sin\delta_{\Km\pip\piz}.
\end{align}
The significant advantage of this analysis over that using two-body final states is that both mixing parameters are measured at first order rather than one at first and one at second order.

The strong phases are not accessible in these measurements but have to come from measurements performed using quantum-correlated \Dz-\Dzb pairs produced at threshold.
Such measurements are available from \cleo~\cite{PhysRevD.73.034024,PhysRevD.77.019901,Rosner:2008fq,Asner:2008ft} and can be further improved at \besiii.

The comparison of effective inverse lifetimes in decays of \Dz (\Dzb) mesons into final states which are \CP eigenstates, $\hat{\Gamma}$ ($\hat{\bar{\Gamma}}$), to that of a Cabibbo-favoured flavour eigenstate ($\Gamma$) leads to the observable
\begin{equation}
\ycp=\frac{\hat{\Gamma}+\hat{\bar{\Gamma}}}{2\Gamma}-1\approx\eta_{\CP}\left[\left( 1 -\frac{A_m^2}{8}\right)y\cos\phi -\frac{A_m}{2}x\sin\phi\right],
\end{equation}
where $A_m$ is the \CP violation in mixing defined alongside the direct \CP violation $A_d$ by $|\lambda_f^{\pm1}|^2\approx(1\pm A_m)(1\pm A_d)$~\cite{Gersabeck:2011xj}.
In the limit of \CP conservation \ycp equals the mixing parameter $y$.
As the \CP-violating contributions $A_m$ and $\phi$ enter only at second order, measurements of \ycp are among the most powerful constraints of the mixing parameter $y$.

Comparing the \CP eigenstates $\Km\Kp$ and $\pim\pip$ to the Cabibbo-favoured mode $\Km\pip$, the \belle~\cite{Staric:2007dt} and \babar~\cite{PhysRevD.80.071103} collaborations have measured $\ycp=(13.1\pm3.2\pm2.5)\times10^{-3}$ and $\ycp=(11.6\pm2.2\pm1.8)\times10^{-3}$, respectively.
These measurements have recently been updated by preliminary results based on the full dataset of flavour-tagged events for both collaborations.
The \babar collaboration has added the larger sample of untagged events to the analysis, however, with limited gain in sensitivity due to larger systematic uncertainties for the untagged sample which has lower purity compared to the \Dstar-tagged events.
The updated results are $\ycp=(7.2\pm1.8\pm1.2)\times10^{-3}$ and $\ycp=(11.1\pm2.2\pm1.1)\times10^{-3}$, for \babar~\cite{Neri:2012} and \belle~\cite{Staric:2012}, respectively.

It is worth noting that the central value of the \babar result is significantly lower compared the one from 2007.
This is thought to be compatible with a statistical fluctuation due to the added data as well as only partial overlap in the older dataset following improvements in reconstruction and analysis.
The new \babar result relaxes the tension that existed between measurements of \ycp, which favoured values of about $1\%$, and other mixing measurements, which tend towards smaller values.
Such a tension would be impossible to be explained by \CP violation as that leads to $\ycp\le y$.

Another possibility of measuring \ycp is using the decay mode \decay{\Dz}{\KS\Km\Kp}.
The \belle collaboration have published a measurement in which they compare the effective lifetime around the \Pphi resonance with that measured in sidebands of the $\Km\Kp$ invariant mass~\cite{Zupanc:2009sy}.
The effective \CP eigenstate content in these regions is determined with two different models.
Their result is $\ycp=(1.1\pm6.1\pm5.2)\times10^{-3}$.

The decay \decay{\Dz}{\KS\Km\Kp} and more so the decay \decay{\Dz}{\KS\pim\pip} give excellent access to the mixing parameters $x$ and $y$ individually.
At the same time they allow a measurement of parameters of indirect \CP violation as discussed in the following section.
Under the assumption of no \CP violation \belle and \babar have measured
\begin{table}[h!!!]
\centering
\begin{tabular}{lcc}
           & $x$ in $10^{-3}$ & $y$ in $10^{-3}$\\
\hline
\belle~\cite{Abe:2007rd} & $8.0 \pm 2.9 \pm 1.7$ & $3.3 \pm 2.4 \pm 1.5$\\
\babar~\cite{delAmoSanchez:2010xz} & $1.6 \pm 2.3 \pm 1.2\pm0.8$ & $5.7 \pm 2.0 \pm 1.3\pm0.7$,
\end{tabular}
\end{table}

\noindent where the last uncertainty in the \babar measurement is a model uncertainty.
The \belle result has recently been updated including the full available dataset based on the final reconstruction~\cite{Tao:2012}.
This leads to $x=(0.56\pm0.19^{+0.03}_{-0.09}\,^{+0.06}_{-0.09})\times10^{-3}$ and $y=(0.30\pm0.15^{+0.04}_{-0.05}\,^{+0.03}_{-0.06})\times10^{-3}$.
Once more, the last uncertainty is a model uncertainty.
Curiously, while the recent \babar result on \ycp has relaxed the tension with $y$, now there is a nearly $3\sigma$ tension among the latest \belle results for \ycp and $y$.
Additional measurements are required to resolve this situation.

The \lhcb collaboration has made its first measurement of \ycp~\cite{Aaij:2011ad} based on two-body \Dz decays recorded in 2010 to $\ycp=(5.5\pm6.3\pm4.1)\times10^{-3}$.
While this measurement falls short of being at the level of precision of the B-factory measurements in this mode, \lhcb has recorded over a factor 50 more in integrated luminosity to date.
The Heavy Flavor Averaging Group (HFAG) has combined all measurements of \ycp~\cite{Amhis:2012hf} to $\ycp=(8.7\pm1.6)\times10^{-3}$.

By the time of this review no single experiment observation of mixing in \Dz mesons with a significance exceeding $5\sigma$ has been possible.
However, the combination of the numerous measurements by HFAG excludes the no-mixing hypothesis by about $10\sigma$~\cite{Amhis:2012hf}.
Under the assumption of no \CP violation in mixing or decays, the world average of the mixing parameters is $x=(6.5^{+1.8}_{-1.9})\times10^{-3}$ and $y=(7.4\pm1.2)\times10^{-3}$.

\section{Charm \CP violation}
\label{sec:cp_charm}

\subsection{Indirect \CP violation}
Indirect \CP violation can be measured through the comparison of effective lifetimes of \Dz and \Dzb decays to \CP eigenstates.
This leads to the observable
\begin{equation}
\agamma=\frac{\hat{\Gamma}-\hat{\bar{\Gamma}}}{\hat{\Gamma}+\hat{\bar{\Gamma}}}\approx\eta_{\CP}\left[\frac{1}{2}\left( A_m+A_d\right)y\cos\phi -x\sin\phi\right],
\label{eqn:agamma}
\end{equation}
which has contributions from both direct and indirect \CP violation~\cite{Gersabeck:2011xj,Kagan:2009gb}.
Currently, there are three measurements of \agamma which are all compatible with zero.
The \belle~\cite{Staric:2007dt}, \babar~\cite{Aubert:2007en} and \lhcb~\cite{Aaij:2011ad} collaborations have measured $\agamma=(0.1\pm3.0\pm1.5)\times10^{-3}$, $\agamma=(2.6\pm3.6\pm0.8)\times10^{-3}$, and $\agamma=(-5.9\pm5.9\pm2.1)\times10^{-3}$, respectively.
With the \lhcb result being based only on a small fraction of the data recorded so far, significant improvements in sensitivity may be expected in the near future.
In parallel to their updates of \ycp, \babar~\cite{Neri:2012} and \belle~\cite{Staric:2012} have also released preliminary results for \agamma based on their full datasets.
They have measured $\agamma=(0.9\pm2.6\pm0.6)\times10^{-3}$ and $\agamma=(-0.3\pm2.0\pm0.8)\times10^{-3}$, respectively.
The HFAG world average~\cite{Amhis:2012hf} yields $\agamma=(-0.2\pm1.6)\times10^{-3}$ in agreement with no \CP violation.

Using current experimental bounds values of \agamma up to $\mathcal{O}(10^{-4})$ are expected from theory~\cite{Kagan:2009gb,Bigi:2011re}.
It has however been shown that enhancements up to about one order of magnitude are possible, for example in the presence of a fourth generation of quarks~\cite{Bobrowski:2010xg} or in a little Higgs model with T-parity~\cite{Bigi:2011re}. 
This would bring \agamma close to the current experimental limits.

Eventually, the interpretation of \CP violation results requires precise knowledge of both mixing and \CP violation parameters.
The relative sensitivity to the \CP-violating quantities in the observable \agamma is limited by the relative uncertainty of the mixing parameters.
Therefore, to establish the nature of a potential non-zero measurement of \agamma it is mandatory to have measured the mixing parameters with a relative precision of $\approx10\%$.

The analyses of the decays \decay{\Dz}{\KS\pim\pip} and \decay{\Dz}{\KS\Km\Kp} offer separate access to the parameters $x$, $y$, $|q/p|$ and $\arg(q/p)$ and are one of the most promising ways of obtaining precise mixing measurements.
These analyses require the determination of the decay-time dependence of the phase space structure (Dalitz plot, see Ref.~\refcite{Dalitz:1953cp}) of these decays.
This can be obtained in two ways: explicit fits of the time evolution of resonances based on Dalitz-plot models, or based on a measurement of the strong-phase difference across the Dalitz-plot carried out by the \cleo collaboration~\cite{Libby:2010eq}.
One measurement made by the \belle collaboration has determined these parameters based on a Dalitz plot model~\cite{Abe:2007rd}.
Other measurements were performed by the \cleo~\cite{Asner:2005sz} and \babar~\cite{delAmoSanchez:2010xz} collaborations assuming \CP conservation and thus extracting only $x$ and $y$.
With the data samples available and being recorded at \lhcb and those expected at future flavour factories, these measurements will be very important to understand charm mixing and \CP violation.
However, in order to avoid systematic limitations it will be important to reduce model uncertainties or to improve model-independent strong-phase difference measurements which are possible at \besiii.

\subsection{Direct \CP violation}
Direct \CP violation is searched for in decay-time integrated measurements.
However, for neutral mesons, the decay-time distribution of the data has to be taken into account to estimate the contribution from indirect \CP violation.
Currently, the most striking measurements have been made in decays of \Dz mesons into two charged pions or kaons.
While early measurements of \babar~\cite{Aubert:2007if} and \belle~\cite{Staric:2008rx} have not shown significant deviations from zero, the \lhcb collaboration has reported first evidence for \CP violation in the charm sector~\cite{Aaij:2011in}.
They have measured 
\begin{displaymath}
\dacp\equiv{}A_{\CP}(\Km\Kp)-A_{\CP}(\pim\pip)=(-8.1\pm2.1\pm1.1)\times10^{-3}.
\end{displaymath}
Meanwhile, the \cdf collaboration has released a preliminary measurement of $\dacp=(-6.2\pm2.1\pm1.0)\times10^{-3}$ which shows a hint of a deviation from zero~\cite{CDF:10784}, in support of the \lhcb result.
Just before submission of this review the \belle collaboration has released an update of their measurement of \dacp based on their full reprocessed dataset showing a roughly $2\sigma$ deviation from zero~\cite{Ko:2012}.

The quantity \dacp exploits first-order cancellation of systematic uncertainties in the difference of asymmetries.
At higher order, terms that are products of individual asymmetries contribute, \eg\ the product of a production and a \CP asymmetry, which no longer cancel.
In a kinematic region with large production asymmetries such a contribution leads to relative corrections of \dacp of the order of the local production asymmetry.
Hence, these higher order terms need to be taken into account for a precision measurement of the size of observed \CP violation.

The observable \dacp gives access to the difference in direct \CP violation of the two decay modes through
\begin{equation}
\dacp=\Delta{}a_{\CP}^{\rm dir}\left(1+\ycp\frac{\overline{\langle{}t\rangle}}{\tau}\right)+\overline{A}_\Gamma\frac{\Delta\langle{}t\rangle}{\tau},
\label{eqn:dacp}
\end{equation}
where $\tau$ is the nominal \Dz lifetime, $\overline{X}\equiv(X(\Km\Kp)+X(\pim\pip))/2$, and $\Delta{}X\equiv{}X(\Km\Kp)-X(\pim\pip)$ for $X=(a_{\CP}^{\rm dir},\langle{}t\rangle)$~\cite{Gersabeck:2011xj}.
Equation~(\ref{eqn:dacp}) assumes the \CP-violating phase $\phi$ to be universal.
For a small non-zero difference in this phase between the two final states, $\Delta\phi_{f}$, an additional term of the form $x\Delta\phi_{f}\overline{\langle{}t\rangle}/\tau$ arises as pointed out in Ref.~\refcite{Kagan:2009gb}.
Given a typical variation of $\overline{\langle{}t\rangle}/\tau$ between $1$ and $2.5$ for the different experiments the contribution of $\Delta\phi_{f}$ is suppressed by $x\overline{\langle{}t\rangle}/\tau\approx10^{-2}$.

While it was commonly stated in literature that \CP violation effects in these channels were not expected to exceed $10^{-3}$, this statement has been revisited in numerous recent publications.
To date, no clear understanding of whether~\cite{Brod:2011re,Feldmann:2012js,Bhattacharya:2012ah,Franco:2012ck} or not~\cite{Bigi:2011re,Rozanov:2011gj,Cheng:2012wr,Li:2012cf} \CP violation of this level can be accommodated within the standard model has emerged.
In parallel to attempts to better the standard model calculations, many estimates of potential effects of physics beyond the standard model have been made~\cite{Bigi:2011re,Feldmann:2012js,Rozanov:2011gj,Grossman:2006jg,Isidori:2011qw,Wang:2011uu,Hochberg:2011ru,Pirtskhalava:2011va,Chang:2012gn,Giudice:2012qq,Altmannshofer:2012ur,Chen:2012am,Gedalia:2012pi,Lodone:2012kp,Brod:2012ud,Hiller:2012wf,Mannel:2012hb,KerenZur:2012fr,Barbieri:2012bh,Bigi:2012ev,Chen:2012us,Dolgov:2012ez,Delaunay:2012cz,Bhattacharya:2012kq}.

Within the standard model the central value can only be explained by significantly enhanced penguin amplitudes.
This enhancement is conceivable when estimating flavour $SU(3)$ or U-spin breaking effects from fits to data of \PD decays into two pseudo scalars~\cite{Feldmann:2012js,Bhattacharya:2012ah,Pirtskhalava:2011va,Brod:2012ud,Golden:1989qx,Cheng:2012xb}.
However, attempts of estimating the long distance penguin contractions directly have failed to yield conclusive results to explain the enhancement.

Lattice QCD has the potential of assessing the penguin enhancement directly.
However, several challenges arise which make these calculations impossible at the moment~\cite{Luscher:1986pf,Luscher:1990ux,Lellouch:2000pv,Blum:2011ng,Blum:2011pu,Hansen:2012tf,Yu:2011gk}.
Following promising results on $\PK\to\pi\pi$ decays, additional challenges arise in the charm sector as $\pi\pi$ and $\PK\PK$ states mix with $\eta\eta$, $4\pi$, $6\pi$ and other states.
Possible methods have been proposed and first results may be expected within the next decade.

General considerations on the possibility of interpreting the $\Delta{}A_{\CP}$ in models beyond the standard model have lead to the conclusion that an enhanced chromomagnetic dipole operator is required.
These operators can be accommodated in minimal supersymmetric models with non-zero left-right up-type squark mixing contributions or, similarly, in warped extra dimensional models~\cite{Grossman:2006jg,Isidori:2011qw,Giudice:2012qq,Randall:1999ee,Goldberger:1999uk,Huber:2000ie,Gherghetta:2000qt}.
Tests of these interpretations beyond the standard model are in the focus of ongoing searches.
One promising group of channels are radiative charm decays where the link between the chromomagnetic and the electromagnetic dipole operator leads to predictions of enhanced \CP asymmetries of several percent~\cite{Isidori:2012yx}.

Another, complementary, test is to search for contributions beyond the standard model in $\Delta I=3/2$ amplitudes.
This class of amplitudes leads to several isospin relations which can be tested in a range of decay modes, e.g.\ $\PD\to\pi\pi$, $\PD\to\rho\pi$, $\PD\to\PK\bar{\PK}$~\cite{Feldmann:2012js,Grossman:2012eb}.
Several of these measurements, such as the Dalitz plot analysis of the decay $\Dz\to\pip\pim\piz$, have been performed by \babar and \belle and will be possible at \lhcb as well as future $\ep\en$ machines.

Beyond charm physics, the chromomagnetic dipole operators would affect the neutron and nuclear EDMs, which are expected to be close to the current experimental bound~\cite{Giudice:2012qq}.
Similarly, rare FCNC top decays are expected to be enhanced, if kinematically allowed.
Furthermore, quark compositness can be related to the $\Delta{}A_{\CP}$ measurement and tested in dijet searches.
Current results favour the new physics contribution to be located in the $\Dz\to\Km\Kp$ decay as the strange quark compositness scale is less well constrained~\cite{Delaunay:2012}.

Another group of channels suitable for \CP violation searches is that of decays of \Dp and \Ds mesons into three charged hadrons, namely pions or kaons.
Here, \CP violation can occur in two-body resonances contributing to these decay amplitudes.
Asymmetries in the Dalitz-plot substructure can be measured using an amplitude model or using model-independent statistical analyses~\cite{Aubert:2008yd,Bediaga:2009tr,Williams:2011cd}.
The latter allow \CP asymmetries to be discovered while eventually a model-dependent analysis is required to identify its source.
The two types of model-independent analyses differ in being either binned~\cite{Aubert:2008yd,Bediaga:2009tr} or unbinned~\cite{Williams:2011cd} in the Dalitz plane.

The binned approach computes a local per-bin asymmetry and judges the presence of \CP violation by the compatibility of the distribution of local asymmetries across the Dalitz plane with a normal distribution.
This method obviously relies on the optimal choice of bins.
Bins ranging across resonances can lead to the cancellation of real asymmetries within a bin.
Too fine binning can reach the limit of statistical sensitivity, whereas too coarse binning can wash out \CP violation effects by combining regions of opposite asymmetry.
A model-inspired choice of binning is clearly useful and this does not create a model-dependence in the way that fitting resonances directly does.
This method does not yield an easy-to-interpret quantitative result.
This issue has been discussed in a recent update of the procedure~\cite{Bediaga:2012tm}.

The unbinned asymmetry search calculates a test statistic that allows the assignment of a $p$-value when comparing to the distribution of the statistic for many random permutations of the events among the particle and anti-particle datasets.
Moreover, being unbinned, there is no need for a model-inspired choice of binning.
The drawback of this method is its requirement on computing power.
The calculation of the test statistic scales as the square of the number of events.

Beyond three-body final states similar analyses can be performed in decays into four hadrons, \eg\ decays of \Dz into four charged hadrons.
This too gives access to interesting resonance structures that may exhibit significant \CP asymmetries.
However, rather than having a two-dimensional Dalitz plane, the phase space for four-body decays is five-dimensional (see \eg\ Ref.~\refcite{Rademacker:2006zx}).
This poses not only a challenge on the visualisation but also on any binned approach due to rapidly decreasing sample sizes per bin.
Also, the phase-space substructure can no longer be described only by interfering amplitudes of pseudo two-body decays as also three-body decays may contribute.
The \lhcb collaboration has released a first model-independent search for \CP violation in the decay \decay{\Dz}{\pim\pip\pip\pim} without finding any hint of \CP non-conservation~\cite{LHCb-CONF-2012-019}.

Neither searches for phase-space integrated asymmetries~\cite{Aitala:1996sh,Link:2000aw,Aubert:2005gj,Alexander:2008aa,Rubin:2008aa,Mendez:2009aa,Aaltonen:2012nd}, nor searches for local asymmetries in the Dalitz plot~\cite{Aubert:2005gj,Rubin:2008aa,Staric:2011en,PhysRevD.84.112008,PhysRevLett.108.071801} have shown any evidence for \CP violation.
The largest signal is the recently reported measurement of \CP violation in \decay{\Dp}{\Pphi\pip} of $A_{\CP}^{\Pphi\pip}=(5.1\pm2.8\pm0.5)\times10^{-3}$ by the \belle collaboration~\cite{PhysRevLett.108.071801}, which exploits cancellation of uncertainties through a comparison of asymmetries in the decays of \Dp and \Ds mesons into the final state $\Pphi\pip$.

Decays of \Dp and \Ds are into a \KS and either a \Kp or a \pip are closely related to their \Dz counterparts.
Measurements of time-integrated asymmetries in these decays are expected to exhibit a contribution from \CP violation in the kaon system.
As pointed out recently~\cite{Grossman:2011to} this contribution depends on the decay-time acceptance of the \KS.
This can lead to different expected values for different experiments.
For \belle~\cite{Ko:2012pe}, the expected level of asymmetry due to \CP violation in the kaon system is $-3.5\times10^{-3}$.
For \lhcb on the other hand, there is no significant asymmetry induced by kaon \CP violation~\cite{Hamish:2012} as the \lhcb acceptance, for \KS reconstructed in the vertex detector, corresponds to about $10\%$ of a \KS lifetime.
\CP violation searches in the decays \decay{\Dp}{\KS\pip}~\cite{PhysRevD.80.071103,Mendez:2009aa,PhysRevLett.104.181602} and \decay{\Ds}{\KS\pip}~\cite{Mendez:2009aa,PhysRevLett.104.181602} show significant asymmetries.
However, these asymmetries are fully accommodated in the expected \CP violation of the kaon system.
These measurements do not show any hint for an asymmetry in \PD decay amplitudes.
Future, more precise, measurements will reveal whether or not these remain in agreement with the expected contribution from the kaon system.

In the light of the recent measurements it is evident that there are four directions to pursue: more precise measurements of \dacp and the individual asymmetries are required to establish the effect; further searches for time-integrated \CP violation need to be carried out in a large range of modes that allow to identify the source of the \CP asymmetry; searches for time-dependent \CP asymmetries, particularly via more precise measurements of \agamma; and finally a more precise determination of the mixing parameters is required.

\subsection{Interplay of mixing, direct and indirect \CP violation}
\begin{figure}
\centering
\includegraphics[width=1.0\textwidth]{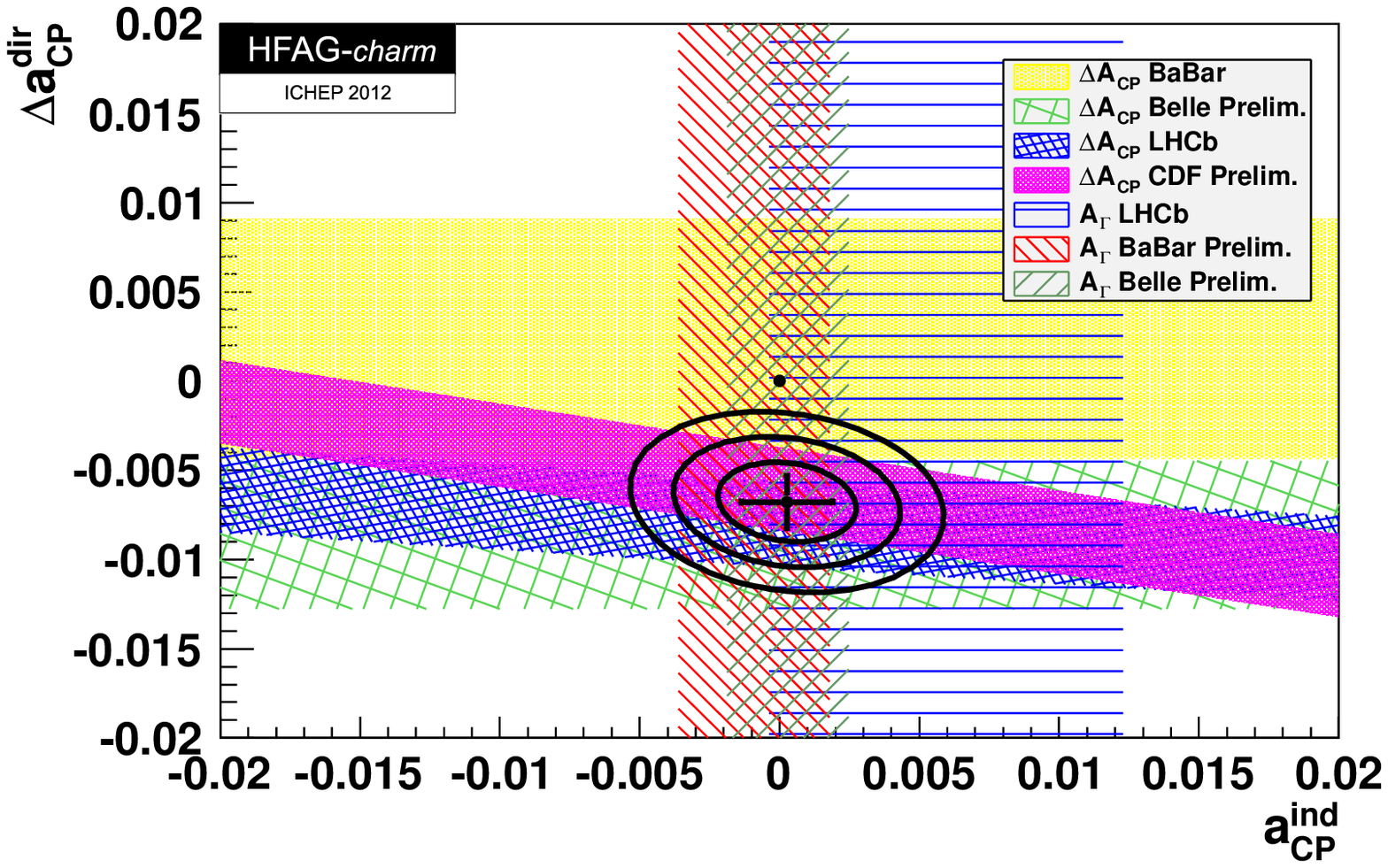}
\caption{Fit of $\Delta{}a_{\CP}^{\rm dir}$ and $a_{\CP}^{\rm ind}$. Reproduced from Ref.~\protect\refcite{Amhis:2012hf}.}
\label{fig:hfag_cpv}
\end{figure}

Following Eqs.~(\ref{eqn:agamma}) and~(\ref{eqn:dacp}) it is obvious that both \agamma and \dacp share the underlying \CP-violating parameters.
Allowing for a non-universal \CP-violating phase $\phi$ one can write
\begin{align}
\agamma(f)&=-a_{\CP}^{\rm ind}-a_{\CP}^{\rm dir}(f)\ycp-x\phi_{f},\label{eq:agamma}\\
A_{\CP}(f)&=a_{\CP}^{\rm dir}(f)-\agamma(f)\frac{\langle{}t\rangle}{\tau},\\
\dacp&=\Delta{}a_{\CP}^{\rm dir}-\Delta\agamma\frac{\overline{\langle{}t\rangle}}{\tau}-\overline{A}_\Gamma\frac{\Delta\langle{}t\rangle}{\tau},
\end{align}
where again $\overline{X}\equiv(X(\Km\Kp)+X(\pim\pip))/2$ and $\Delta{}X\equiv{}X(\Km\Kp)-X(\pim\pip)$ for $X=(a_{\CP}^{\rm dir},\langle{}t\rangle)$.
It is expected that, at least within the standard model, one has $a_{\CP}^{\rm dir}(\Km\Kp)=-a_{\CP}^{\rm dir}(\pim\pip)$ and thus $\overline{A}_\Gamma=-a_{\CP}^{\rm ind}$.
This set of equations shows that it is essential to measure both time-dependent (\agamma) and time-integrated asymmetries ($A_{\CP}$) separately in the decay modes \decay{\Dz}{\Km\Kp} and \decay{\Dz}{\pim\pip} in order to distinguish the various possible sources of \CP violation.
Currently, the experimental precision on \agamma is such that there is no sensitivity to differences in the contributions from direct \CP violation to measurements using $\Km\!\Kp$ or $\pim\!\pip$ final states.
Hence, the approximation $\agamma\equiv\overline{A}_\Gamma\approx\agamma(\Km\Kp)\approx\agamma(\pim\pip)$ can be used to obtain
\begin{align}
\agamma&=-a_{\CP}^{\rm ind}\\
\dacp&=\Delta{}a_{\CP}^{\rm dir}\left(1+\ycp\frac{\overline{\langle{}t\rangle}}{\tau}\right)-a_{\CP}^{\rm ind}\frac{\Delta\langle{}t\rangle}{\tau}.
\end{align}
These equations have been used by HFAG to prepare a fit of the direct and indirect \CP violation contributions~\cite{Amhis:2012hf} as shown in Fig.~\ref{fig:hfag_cpv}.
This fit yields a confidence level of about $2\times10^{-5}$ for the no \CP violation hypothesis with best fit values of $\Delta{}a_{\CP}^{\rm dir}=(-6.78\pm1.47)\times10^{-3}$ and $a_{\CP}^{\rm ind}=(0.27\pm1.63)\times10^{-3}$.
This also shows that the most likely source of the large measured values for \dacp is direct \CP violation in one or both of the relevant decay modes.
The fit formalism will have to be refined using the equations discussed above in the future as more precise measurements as well as individual asymmetries will become available.

The ultimate goal of mixing and \CP violation measurements in the charm sector is to reach precisions at or below the standard model predictions.
In some cases this requires measurements in several decay modes in order to distinguish enhanced contributions of higher order standard model diagrams from effects caused by new particles.

Indirect \CP violation measurements at \lhcb are mostly constrained by the observable \agamma (see Eq.~(\ref{eq:agamma})).
The \CP violating parameters in this observable are multiplied by the mixing parameters $x$ and $y$, respectively.
Hence, the relative precision on the \CP violating parameters is limited by the relative precision of the mixing parameters.
Therefore, aiming at a relative precision below $10\%$ for the \CP violation quantities and taking into account the current mixing parameter world averages, the target precision for the mixing parameters is $2-3\times10^{-4}$.
With standard model indirect \CP violation expected of the order of $10^{-4}$, the direct \CP violation parameter contributing to \agamma has to be measured to an absolute precision of $10^{-3}$ in order to distinguish the two types of \CP violation in \agamma.

Direct \CP violation is not expected to be as large as the current world average of $\Delta A_{\CP}$ in other decay modes.
Therefore a precision of $5\times10^{-4}$ or better for asymmetry differences as well as individual asymmetries is needed for measurements of other singly-Cabibbo-suppressed charm decays.
While measurements of time-integrated raw asymmetries at this level should be well within reach, the challenge lies in the control of production and detection asymmetries in order to extract the physics asymmetries of individual decay modes.

For multibody final states the aim is clearly the understanding of \CP asymmetries in the interfering resonances rather than global asymmetries.
Of highest interest are those resonances that are closely related to the two-body modes used in $\Delta A_{\CP}$, for example the vector-pseudoscalar resonances $\Kstar\PK$ and $\Prho\Ppi$.
The measurement of further suppressed resonances is of interest as well since those have no contributions from gluonic penguin diagrams, thus allowing to constrain the source of \CP violating effects.

\section{Rare charm decays}
Rare decays provide a wide range of interesting measurements.
The list of decay modes includes flavour-changing neutral currents, radiative, lepton-flavour violation, lepton-number violation, as well as baryon-number violation.
While a full discussion of rare charm decays would be beyond the scope of this review a few remarks shall be made here.

There is a direct link between mixing and flavour changing neutral current decays in several extensions of the standard model~\cite{Golowich:2007ka,Golowich:2009ii}.
These relate $\Delta C=1$ annihilation amplitudes to $\Delta C=2$ mixing amplitudes where the annihilation product creates a new \CP-conjugated $\cquark\ubarquark$ pair.
At tree level one example is a heavy $Z$-like boson with non-zero flavour-changing couplings.

The current central values of the \Dz-mixing parameters translate into model-dependent limits for rare decays based on common amplitudes.
These rare decay limits lie significantly below the current experimental limits.
As \Dz mixing is well established any upper limit from mixing will not change significantly in the future.
However, due to the direct correlation of mixing and rare decays, any observation above the model-dependent rare decay limits will rule out the corresponding model.
The best limit on flavour-changing neutral current decays is the recent \lhcb limit on the decay \decay{\Dz}{\mun\mup} of $1.1\times{}10^{-8}$ at $90\%$ confidence level~\cite{LHCb-CONF-2012-005}.

Among lepton-flavour violating decays the most stringent constraint is a \belle search for \decay{\Dz}{\mump\epm} achieving a limit of $2.6\times{}10^{-7}$ at $90\%$ confidence level~\cite{Petric:2010yt}.
Searches for lepton-flavour violating muon or kaon decays already provide more constraining limits, however, in scenarious of non-universal couplings charm decays, giving access to the up-quark sector, are of great interest.

The best limit on lepton-number violating charm decays has been placed by \babar on the decay \decay{\Dp}{\Km\ep\ep} with a limit of $9.0\times{}10^{-7}$ at $90\%$ confidence level~\cite{Lees:2011hb}.
Only the \cleo collaboration has carried out searches for baryon-number violating charm decays.
Their best limit on the decay \decay{\Dz}{p\en} is $10^{-5}$ at $90\%$ confidence level~\cite{Rubin:2009aa}.
For a more complete overview of rare charm decays please refer to Ref.~\cite{Amhis:2012hf}.

\section{Conclusion}
The first evidence for \CP violation in the charm sector has opened the door wide for a broad range of measurements.
40 years after the observation of the first hint of charm particles in cosmic rays and exactly 37 years after the dicovery of \Jpsi mesons, charm measurements may have shown first hints of effects beyond the standard model at the LHC.
The unambiguous observation of \CP violation will be the near term goal.
The ultimate task is the interpretation of the observed effects for which theoretical and experimental communities have to collaborate closely to overcome the hurdles related to the charm quark mass and the large cancellations in this system.

With the \PB factories, \cleo-c and \cdf analysing their final datasets, most new results are expected to come from \lhcb and BESIII.
These are expected to explore very interesting territory for charm \CP violation.
The longer term future will be shaped by the \lhcb upgrade as well as future $\ep\en$ collider experiments running both at the beauty and charm thresholds.
Charm's third time has just begun to yield first fruits which may well develop into a real charm.

\section*{Acknowledgments}
The author would like to thank Yuval Grossman, Claus Grupen, Alex Kagan, Alexander Lenz, and Alexey Petrov as well as the members of the LHCb collaboration for very insightful discussions.
Special thanks to Alexander Lenz and Johannes Albrecht who dared to read the article before submission.
The author further acknowledges the support of a Marie Curie Action: ``Cofunding of the CERN Fellowship Programme (COFUND-CERN)'' of the European Community's Seventh Framework Programme under contract number (PCOFUND-GA-2008-229600).

\bibliographystyle{utphys}       
\bibliography{charm}   

\end{document}